\documentclass[pre, twocolumn, showkeys, showpacs, amsmath, amssymb]{revtex4}

\usepackage{graphicx}
\begin{document}

\title{Selective advantage for sexual reproduction with random haploid fusion}

\author{Emmanuel Tannenbaum}
\affiliation{Department of Chemistry, Ben-Gurion University of the Negev, Be'er-Sheva, Israel}
\email{emanuelt@bgu.ac.il}

\begin{abstract}

This paper develops a simplified set of models describing asexual and sexual replication in unicellular diploid organisms.  The models assume organisms whose genomes consist of two chromosomes, where each chromosome is assumed to be functional if it is equal to some master sequence $ \sigma_0 $, and non-functional otherwise.  The first-order growth rate constant, or fitness, of an organism, is determined by whether it has zero, one, or two functional chromosomes in its genome.  We assume that an organism with no functional chromosomes has zero fitness.  For a population replicating asexually, a given cell replicates both of its chromosomes, and then the cell divides and splits its genetic material evenly between the two cells.  For a population replicating sexually, a given cell first divides into two haploids, which enter a haploid pool.  Within the haploid pool, haploids fuse into diploids, which then divide via the normal mitotic process.  Haploid fusion is modeled as a second-order rate process.  We review the previously studied case of selective mating, where it is assumed that only haploids with functional chromosomes can fuse, and also consider the case of random haploid fusion.  When the cost for sex is small, as measured by the ratio of the characteristic haploid fusion time to the characteristic growth time, we find that sexual replication with random haploid fusion leads to a greater mean fitness for the population than a purely asexual strategy.  However, independently of the cost for sex, we find that sexual replication with a selective mating strategy leads to a higher mean fitness than the random mating strategy.  This result is based on the assumption that a selective mating strategy does not have any additional time or energy costs over the random mating strategy, an assumption that is discussed in the paper.  The results of this paper are consistent with previous studies suggesting that sex is favored at intermediate mutation rates, for slowly replicating organisms, and at high population densities.  

\end{abstract}


\maketitle

\section{Introduction}

The evolution and maintenance of sexual replication is one of the central questions in modern evolutionary biology \cite{Bell:82, Williams:75, Smith:78, Michod:95, Hurst:96}.  The theories with the broadest acceptance are that sex allows for the removal of deleterious mutations from a population \cite{Michod:95, Bernstein:85, Muller:64}, or that sex allows for faster adaptation in dynamic environments \cite{Bell:82, Hamilton:90}.

As has been noted in previous studies \cite{Tannenbaum:06, TannFon:06}, the various theories for the existence of sex are incomplete, in that they do not explain why some organisms are obligately sexual, while other organisms either alternate between asexual and sexual replication, or are asexual replicators with some ability for recombination with other organisms.  

In a recent set of papers \cite{Tannenbaum:06, TannFon:06, LeeTann:07, Tannenbaum:07}, Tannenbaum, Fontanari, and Lee studied the competition between asexual and sexual replicators in various regimes.  Two of the models consider single-celled organisms, that replicate by dividing into haploids.  These haploids then enter a haploid pool, where they fuse with other haploids.  The resulting diploid then divides through the normal mitotic process.  

Two other models were developed to be more appropriate for modeling multicellular organisms, which release either asexual spores or gametes.  Here, we considered the case where the population produces identical gametes, and where the population produces distinct sperm and egg gametes.

For both sets of models it was found that sexual replication is favored in slowly replicating organisms and high population densities.  For the case of multicellular organisms, it was found that distinct sperm and egg gametes were necessary to maintain a selective advantage for sexual replication over asexual replication \cite{Tannenbaum:07}.

In all the models considered, the authors assumed that there is a time cost for sex, due to the time it takes for a given haploid to find another haploid with which to fuse.  When the time cost for sex is low, then the selective advantage for sex outweighs the fitness penalty.  Once the time cost for sex becomes sufficiently large, the selective advantage for sex no longer outweighs the fitness penalty, and asexual replication becomes the preferred replication strategy.

In order to facilitate an exact analysis of the models, the authors made a number of simplifiying assumptions:  It was assumed that the organisms have diploid genomes consisting of only two chromosomes, and that each chromosome is only considered functional, or viable, if it is equal to some master sequence.  It was also assumed that the fitness of the organisms is determined by the number of viable chromosomes in the genome ($ 0 $, $ 1 $, $ 2 $), and that organisms with two defective chromosomes have zero fitness. 
   
Finally, one of the key simplifications that was made in the sexual replication models was the assumption that only haploids with viable chromosomes can fuse with one another.  The reasoning behind this selective mating strategy was that haploids with defective chromosomes are not viable, and simply cannot participate further in the replication process.

While the assumption of a selective mating strategy is the easiest one to study, and while it is broadly consistent with the observation that organisms do engage in mate selection, it is nevertheless an overly restrictive assumption that must be relaxed if one wants to develop more realistic sexual replication models.  The reason for this is that it is impossible, in practice, for one organism to read another organism's genome.  Even if this were possible, the correlation between the genome sequence and fitness is extremely difficult to obtain.  

Organisms that engage in mate selection look for certain behaviors and physical attributes, known as {\it indicator traits}, that suggest that the given organism has a good genome \cite{Andersson:94}.  However, the map from such traits to genome is not exact.  Furthermore, the more accuracy one wants in assessing the fitness of an organism, the more time one has to spend studying the organism.  For, if assessing the fitness of an organism is equivalent to reading its genome, then the more accuracy one wants, the more of the genome that has to be read.

Therefore, a selective mating strategy has an additional time cost over other mating strategies, leading to a fitness penalty that may eliminate the advantage for sexual replication entirely.  This time cost was not explicitly considered in previous studies.

Even if the time cost for a selective mating strategy were explicitly considered, we have shown that there will always be an uncertainly as to the exact fitness of the organism.  As a result, more realistic models for sexual replication need to consider mating strategies that take into account the uncertainty that one organism has of another's fitness.  Otherwise, the conclusions that we have drawn regarding the selective advantage for sex are open to the criticism that they are based on an unrealistic and overly restrictive mating strategy.

In this paper, we take the opposite extreme from a selective mating strategy, and consider sexual replication with a random mating strategy.  That is, we assume that all haploids participate in the replication process, and fuse with one another at random.    As long as the cost for sex is negligible, we find that, even with this non-selective mating strategy, the mean fitness for a sexually replicating population is greater than that of the corresponding asexually replicating population.  Nevertheless, if the time cost associated with the selective mating strategy is negligible as well, then the mean fitness of the selective mating strategy is greater than that of the random mating strategy, independent of the cost for sex.  As the cost for sex increases, sexual replication via either the selective or random mating strategies only outcompetes asexual replication over progressively smaller ranges of replication fidelities, and ceases to be advantageous entirely once the cost for sex crosses a threshold value.

This paper is organized as follows:  In the following section (Section II), we develop and analyze the mutation-selection equations appropriate for analyzing an asexually replicating, two-chromosomed, diploid population.  In Section III, we develop the mutation-selection equations appropriate for analyzing the sexually replicating analogue of the population considered in Section II.  We consider both selective and random mating strategies.  Although the selective mating strategy has already been studied, we review it here for the sake of completeness and consistency of notation.  In Section IV, we compare all three replication strategies, and show that a random mating strategy outcompetes asexual replication when the cost for sex is negligible.  We also show that when the cost of implementing the selective mating strategy is negligible, then the selective mating strategy outcompetes the random mating strategy, independently of the cost for sex.  Section V is the Discussion, where we review and discuss the implications of our results.  Finally, in Section VI we conclude the paper with a summary of the main results of the paper, as well as our plans for future research.

\section{The Asexual Replication Model}

We consider a population of unicellular organisms whose genomes consist of two chromosomes.  We assume that a given chromosome, denoted $ \sigma $, is functional if and only if it is equal to some ``master", or wild-type, chromosome $ \sigma_0 $.  Assuming first-order exponential growth, we then assume that the fitness, or first-order growth rate constant, of a given genome is determined by whether the genome consists of zero, one, or two functional chromosomes.  To this end, we let $ \kappa_{vv} $ denote the first-order growth rate constant of organisms with two functional, or equivalently, viable chromosomes.  We let $ \kappa_{vu} $ denote the first-order growth rate constant of organisms with one functional and one non-functional (unviable) chromosome.  Finally, we let $ \kappa_{uu} $ denote the first-order growth rate constant of organisms with two non-functional chromosomes.  

We assume that $ \kappa_{vv} \geq \kappa_{vu} \geq \kappa_{uu} $. We will also assume that $ \kappa_{uu} = 0 $, which makes sense, since an organism with two defective chromosomes is not expected to grow.  We also define $ \kappa = \kappa_{vv} $, and $ \alpha = \kappa_{vu}/\kappa_{vv} $. 

We now divide the populations into three subpopulations:  We let $ n_{vv} $, $ n_{vu} $, and $ n_{uu} $ denote the number of organisms with two, one, and zero viable chromosomes.  We let $ n $ be the total number of organisms, so that $ n = n_{vv} + n_{vu} + n_{uu} $.  The population fractions $ x_{vv} $, $ x_{vu} $, and $ x_{uu} $ are then defined via $ x_{vv} = n_{vv}/n $, $ x_{vu} = n_{vu}/n $, and $ x_{uu} = n_{uu}/n $.

To develop the mutation-selection equations describing the evolutionary dynamics of the asexually replicating equations, we assume that the replication of each cell occurs as follows:  The two chromosomes line up along the center of the cell and replicate (see Figure 1).  Because replication is in general not error-free, we let $ p $ denote the probability that a given genome is replicated correctly.    If the genome is sufficiently long, then the probability that a mutation will occur in a previously mutated region of the genome is negligible, so that an unviable chromosome produces an unviable daughter with probability $ 1 $.  This assumption is known as the {\it neglect of backmutations}.  

We also let $ s $ denote a co-segregation parameter, which is simply the probability that a parent chromosome co-segregates with the other parent chromosome in the cell.  Figure 1 illustrates the various parent cell configurations and the final daughter cell configurations, along with their associated probabilities.  It should be noted that $ 1 - s $ is the same as the $ r $ parameter defined in \cite{TannFon:06}.

With these definitions in hand, we may develop expressions for $ d n_{vv}/dt $, $ d n_{vu}/dt $, and $ d n_{uu}/dt $.  Changing variables from population numbers to population fractions, we obtain the mutation-selection equations,

\begin{eqnarray}
&  &
\frac{d x_{vv}}{dt} = [(2 A_{vv} - 1) \kappa_{vv} - \bar{\kappa}(t)] x_{vv}
\nonumber \\
&  &
\frac{d x_{vu}}{dt} = [\kappa_{vu} p - \bar{\kappa}(t)] x_{vu} + 2 B_{vv} \kappa_{vv} x_{vv}
\nonumber \\
&  &
\frac{d x_{uu}}{dt} = [\kappa_{uu} - \bar{\kappa}(t)] x_{uu} + (1 - p) \kappa_{vu} x_{vu} 
\nonumber \\
&  &
+ s (1 - p)^2 \kappa_{vv} x_{vv}
\nonumber \\
\end{eqnarray}
where $ \bar{\kappa}(t) \equiv (1/n) (dn/dt) = \kappa_{vv} x_{vv} + \kappa_{vu} x_{vu} + \kappa_{uu} x_{uu} $, and,
\begin{eqnarray}
&  &
A_{vv} = \frac{1}{2}[2 p + s (1 - p)^2]
\nonumber \\
&  &
B_{vv} = [1 - p] [1 - s (1 - p)]
\end{eqnarray}

The quantity $ \bar{\kappa}(t) $ is the mean fitness of the population, since it measures the first-order growth rate of the population as a whole.  In order to determine which replication strategy is advantageous in a given regime, we compare the steady-state mean fitnesses of the populations employing the various strategies.  The population with the largest mean fitness will drive the others to extinction, and so the corresponding replication strategy is the advantageous one.

From quasispecies theory, it may be shown that the above system of equations converges to a steady-state, and that the steady-state mean fitness is given by $ \bar{\kappa}(t = \infty) = \max \{(2 A_{vv} - 1) \kappa_{vv}, \kappa_{vu} p \} $, assuming that $ \kappa_{uu} = 0 $ \cite{Alves:97}.  This implies that there exists a $ p_{crit} \in [0, 1] $ such that $ \bar{\kappa}(t = \infty) = 2 A_{vv} - 1 $ for $ p \in [p_{crit}, 1] $, and $ \bar{\kappa}(t = \infty) = \kappa_{vu} p $ for $ p \in [0, p_{crit}] $.

If we define $ \phi_{a} = \bar{\kappa}(t = \infty)/\kappa_{vv} $, then we have that $ \phi_{a} = 2 A_{vv} - 1 $ for $ p \in [p_{crit}, 1] $, and $ \alpha p $ for $ p \in [0, p_{crit}] $, where $ p_{crit} $ is defined by the equality $ \alpha p = 2 A_{vv} - 1 $.

Note that $ \phi_a $ is simply the steady-state mean fitness of the asexual population, normalized by the fitness of the wild-type, $ \kappa_{vv} $.  When we analyze the sexual replication models, we will also work with the normalized mean fitnesses, as it will prove convenient to do so.  When comparing the various replication strategies, the strategy with the largest normalized mean fitness at steady-state is the one that will outcompete the others for the given set of parameters.  

\section{The Sexual Replication Models}

\subsection{The general model}

The general sexual replication model we are considering is illustrated in Figure 2:  A diploid cell grows to mature size, with a first-order growth rate constant given by $ \kappa_{vv} $, $ \kappa_{vu} $, or $ \kappa_{uu} = 0 $, depending on whether the genome has two, one, or zero functional chromosomes, respectively.  The mature diploid then divides into two haploids, which enter a haploid pool.  The haploids fuse with one another, and the resulting diploids then immediately divide via the normal mitotic process.

It should be noted that the corresponding first-order growth rate constants for the asexual and sexual populations are taken to be equal, since the first-order growth rate constants measure the characteristic time it takes a diploid to double in size before dividing.

With sexual replication, it is necessary to keep track of the haploid as well as the diploid population.  Therefore, in addition to the quantities $ n_{vv} $, $ n_{vu} $, and $ n_{uu} $ defined in the previous section, we also have the quantities $ n_v $ and $ n_u $, corresponding to the number of viable and unviable haploids, respectively.

The haploid fusion process is modeled as a binary collision reaction characterized by second-order rate constants $ \gamma_{vv} $, $ \gamma_{vu} $, and $ \gamma_{uu} $, corresponding to the $ v-v $, $ v-u $, and $ u-u $ haploid collisions.  It should be noted that $ \gamma_{vv} $ and $ \gamma_{uu} $ are defined so that $ (\gamma_{vv}/V) n_v^2 $ and $ (\gamma_{uu}/V) n_u^2 $ are the rates of disappearance of the $ v $ haploids and $ u $ haploids respectively, due to $ v-v $ and $ u-u $ haploid fusion respectively.  The quantity $ \gamma_{vu} $ is defined so that $ (\gamma_{vu}/V) n_v n_u $ is the rate of disappearance of either the $ v $ haploids or the $ u $ haploids due to $ v-u $ haploid fusion.

We also assume that the system volume increases so as to maintain a constant density of genomes in the population.  That is, if $ V $ denotes the volume of the system, then we assume that $ \rho \equiv [n_{vv} + n_{vu} + n_{uu} + (1/2) (n_v + n_u)]/V $ is constant.

We now make the following definitions:  We define $ n = n_{vv} + n_{vu} + n_{uu} $ to be the total population of diploids, and $ \bar{\kappa}(t) = (1/n) (dn/dt) $ to be the mean fitness of the diploid population.  We define the population ratios $ x_{vv} = n_{vv}/n $, $ x_{vu} = n_{vu}/n $, $ x_{uu} = n_{uu}/n $, $ x_v = n_v/n $ and $ x_u = n_u/n $.  Finally, we define a diploid density $ \rho^{*} = 
n/V = \rho/(1 + (1/2) (x_v + x_u) $.

If we write down the differential equations governing the values of $ n_{vv} $, $ n_{vu} $, $ n_{uu} $, $ n_v $ and $ n_u $, then changing variables to the population ratios gives the mutation-selection equations,

\begin{eqnarray}
&  &
\frac{d x_{vv}}{dt} = -(\kappa_{vv} + \bar{\kappa}(t)) x_{vv} + \gamma_{vv} \rho^{*} A_{vv} x_v^2
\nonumber \\
&  &
\frac{d x_{vu}}{dt} = -(\kappa_{vu} + \bar{\kappa}(t)) x_{vu} + \gamma_{vv} \rho^{*} B_{vv} x_v^2
\nonumber \\
&  &
+ \gamma_{vu} \rho^{*} B_{vu} x_v x_u
\nonumber \\
&  &
\frac{d x_{uu}}{dt} = -(\kappa_{uu} + \bar{\kappa}(t)) x_{uu} + \gamma_{vv} \rho^{*} C_{vv} x_v^2  
\nonumber \\
&  &
+ \gamma_{vu} \rho^{*} C_{vu} x_v x_u 
+ \gamma_{uu} \rho^{*} x_u^2
\nonumber \\
&  &
\frac{d x_v}{dt} = -\bar{\kappa}(t) x_v + 2 \kappa_{vv} x_{vv} + \kappa_{vu} x_{vu}
\nonumber \\
&  &
- \gamma_{vv} \rho^{*} x_v^2 - \gamma_{vu} \rho^{*} x_v x_u
\nonumber \\
&  &
\frac{d x_u}{dt} = -\bar{\kappa}(t) x_u + \kappa_{vu} x_{vu} + 2 \kappa_{uu} x_{uu} 
\nonumber \\
&  &
- \gamma_{vu} \rho^{*} x_v x_u 
- \gamma_{uu} \rho^{*} x_u^2
\end{eqnarray}
where
\begin{eqnarray}
\bar{\kappa}(t) 
& = &
-\kappa_{vv} x_{vv} - \kappa_{vu} x_{vu} - \kappa_{uu} x_{uu} 
\nonumber \\
&  &
+ \rho^{*} (\gamma_{vv} x_v^2 + 2 \gamma_{vu} x_v x_u + \gamma_{uu} x_u^2)
\end{eqnarray}
and where we have defined the additional quantities $ B_{vu} $, $ C_{vv} $, and $ C_{vu} $ via,
\begin{eqnarray}
&  &
B_{vu} = 1 + p
\nonumber \\
&  &
C_{vv} = \frac{1}{2} s (1 - p)^2
\nonumber \\
&  &
C_{vu} = 1 - p
\end{eqnarray}

Although we are defining the mean fitness of the population with respect to the diploid organisms, at steady-state any two mean fitnesses defined with respect to two distinct sets of subpopulations will be equal.  The reason for this is that, at steady-state, the population reaches a mutation-selection balance, so that any two given sets of subpopulations will be in a fixed proportion to one another (e.g. the total diploid population versus the total population).  Therefore, the per capita rate of increase of one subpopulation is equal to the per capita rate of increase of another.

As a result, there is no ambiguity as to which sexual mean fitness to use when comparing whether a given sexual population outcompetes another sexual population, or an asexual population.  

\subsection{Steady-state for the selective mating strategy}

The selective mating strategy is defined by $ \gamma_{vv} = \gamma $, $ \gamma_{vu} = \gamma_{uu} = 0 $.  This implies that only the $ v $ haploids are allowed to mate, while the $ u $ haploids are essentially thrown away.  One justification for this mating strategy is that the $ u $ haploids contain defective genomes, and so are simply physically incapable of participating further in the replication process.  Another justification, one that is likely more relevant to actual organisms, is that the $ v $ haploids have a way of determining the fitness of a potential haploid mate (via ``indicator" traits, for example), and choose to only mate with the fittest haploids.

For the selective mating strategy, the steady-state equations are given by,
\begin{eqnarray}
&  &
0 = -(\kappa_{vv} + \bar{\kappa}(t = \infty)) x_{vv} + \gamma \rho^{*} x_v^2 A_{vv}
\nonumber \\
&  &
0 = -(\kappa_{vu} + \bar{\kappa}(t = \infty)) x_{vu} + \gamma \rho^{*} x_v^2 B_{vv}
\nonumber \\
&  &
0 = -\bar{\kappa}(t = \infty) x_v + 2 \kappa_{vv} x_{vv} + \kappa_{vu} x_{vu} - \gamma \rho^{*} x_v^2
\nonumber \\
&  &
0 = -\bar{\kappa}(t = \infty) x_u + \kappa_{vu} x_{vu}
\end{eqnarray}
where $ \bar{\kappa}(t = \infty) = -\kappa_{vv} x_{vv} - \kappa_{vu} x_{vu} + \gamma \rho^{*} x_v^2 $.  We purposely neglect the steady-state equation corresponding to $ d x_{uu}/dt = 0 $, since this equation will not be necessary to determine the mean fitness of the population.

Solving the last equation for $ \gamma \rho^{*} x_v^2 $ and substituting the result into the expression for $ \bar{\kappa}(t = \infty) $ gives,
\begin{equation}
\bar{\kappa}(t = \infty) = \kappa_{vv} x_{vv} - \bar{\kappa}(t = \infty) x_v
\end{equation}
and so,
\begin{equation}
x_v = \frac{\kappa_{vv} x_{vv}}{\bar{\kappa}(t = \infty)} - 1
\end{equation}

We also have,
\begin{equation}
x_u = \frac{\kappa_{vu} x_{vu}}{\bar{\kappa}(t = \infty)}
\end{equation}

Substituting these values into the expression for $ \bar{\kappa}(t = \infty) $, and making use of the fact that $ \rho^{*} = \rho/(1 + (1/2) (x_v + x_u)) $ gives, after some manipulation,
\begin{equation}
\frac{(x_{vv} - \phi_{ss})^2}{\phi_{ss} (x_{vv} + \alpha x_{vu} + \phi)^2} = \frac{1}{2} \frac{\kappa_{vv}}{\gamma \rho}
\end{equation}
where $ \phi_{ss} $ is the normalized steady-state mean fitness of the population, and is given by $ \phi_{ss} = \bar{\kappa}(t = \infty)/\kappa_{vv} $.  Here, $ ss $ stands for ``selective sexual".

The steady-state equations for $ x_{vv} $ and $ x_{vu} $ then give,
\begin{eqnarray}
&  &
(1 + \phi_{ss}) x_{vv} = A_{vv} (x_{vv} + \alpha x_{vu} + \phi_{ss})
\nonumber \\
&  &
(\alpha + \phi_{ss}) x_{vu} = B_{vv} (x_{vv} + \alpha x_{vu} + \phi_{ss})
\end{eqnarray}

Therefore, $ x_{vu}/x_{vv} = B_{vv}/A_{vv} \times (1 + \phi_{ss})/(\alpha + \phi_{ss}) $, so that
\begin{eqnarray}
&  &
(1 + \phi_{ss}) x_{vv} = A_{vv} \phi_{ss} + (A_{vv} + (1 + \phi_{ss}) B_{vv} \frac{\alpha}{\alpha + \phi_{ss}}) x_{vv}
\nonumber \\
&  &
\Rightarrow
x_{vv} = \frac{A_{vv} \phi_{ss} (\alpha + \phi_{ss})}{(1 + \phi_{ss}) (\alpha + \phi_{ss} - B_{vv} \alpha) - A_{vv} (\alpha + \phi_{ss})}
\end{eqnarray}

Now, the first steady-state equation for $ x_{vv} $ may be solved to give, $ x_{vv} + \alpha x_{vu} + \phi_{ss} = (1 + \phi_{ss}) x_{vv}/A_{vv} $, which may be substituted into Eq. (10) to give,
\begin{equation}
\frac{A_{vv}^2 (1 - \frac{\phi_{ss}}{x_{vv}})^2}{\phi_{ss} (1 + \phi_{ss})^2} = \frac{1}{2} \frac{\kappa_{vv}}{\gamma \rho} 
\end{equation}

Substituting in the value for $ x_{vv} $ gives, after some manipulation, that,
\begin{equation}
\frac{[\phi_{ss}^2 - (2 A_{vv} (1 - \alpha) - 1 + \alpha p) \phi_{ss} - \alpha p]^2}{\phi_{ss} (\phi_{ss} + 1)^2 (\phi_{ss} + \alpha)^2} = \frac{1}{2} \frac{\kappa_{vv}}{\gamma \rho}
\end{equation}
 
We should note that the results we have obtained here for the selective mating strategy were previously derived by Tannenbaum and Fontanari \cite{TannFon:06}.  Nevertheless, we reviewed them here for the sake of completeness.
 
\subsection{Steady-state for the random mating strategy}

The random mating strategy is defined by $ \gamma_{vv} = \gamma_{vu} = \gamma_{uu} = \gamma $.  Here, any haploid pair is equally likely to fuse as any another haploid pair.

The steady-state equations are,
\begin{eqnarray}
&  &
0 = -(\kappa_{vv} + \bar{\kappa}(t = \infty)) x_{vv} + \gamma \rho^{*} x_v^2 A_{vv}
\nonumber \\
&  &
0 = -(\kappa_{vu} + \bar{\kappa}(t = \infty)) x_{vu} + \gamma \rho^{*} x_v^2 B_{vv} + \gamma \rho^{*} x_v x_u B_{vu}
\nonumber \\
&  &
0 = -\bar{\kappa}(t = \infty) (x_v + x_u) + 2 (\kappa_{vv} x_{vv} + \kappa_{vu} x_{vu})
\nonumber \\
&  &
- \gamma \rho^{*} (x_v + x_u)^2
\nonumber \\
&  &
0 = -\bar{\kappa}(t = \infty) x_v + 2 \kappa_{vv} x_{vv} + \kappa_{vu} x_{vu} - \gamma \rho^{*} x_v (x_v + x_u)
\nonumber \\
\end{eqnarray}
where $ \bar{\kappa}(t = \infty) = -\kappa_{vv} x_{vv} - \kappa_{vu} x_{vu} + \gamma \rho^{*} (x_v + x_u)^2 $.  The third equation is obtained by adding the equations corresponding to $ d x_v/dt = d x_u/dt = 0 $.

We then have,
\begin{equation}
\bar{\kappa}(t = \infty) = \kappa_{vv} x_{vv} + \kappa_{vu} x_{vu} - \bar{\kappa}(t = \infty) (x_v + x_u)
\end{equation}
so that,
\begin{equation}
x_v + x_u = \frac{\kappa_{vv} x_{vv} + \kappa_{vu} x_{vu}}{\bar{\kappa}(t = \infty)} - 1
\end{equation}

If we define $ \tilde{x}_v = x_v/(x_v + x_u) $, $ \tilde{x}_u = x_u/(x_v + x_u) $, then,
\begin{equation}
0 = 2 \kappa_{vv} x_{vv} + \kappa_{vu} x_{vu} - 2 \tilde{x}_v (\kappa_{vv} x_{vv} + \kappa_{vu} x_{vu})
\end{equation}
and so,
\begin{eqnarray}
&  &
\tilde{x}_v = \frac{2 \kappa_{vv} x_{vv} + \kappa_{vu} x_{vu}}{2 (\kappa_{vv} x_{vv} + \kappa_{vu} x_{vu})}
\nonumber \\
&  &
\tilde{x}_u = \frac{\kappa_{vu} x_{vu}}{2 (\kappa_{vv} x_{vv} + \kappa_{vu} x_{vu})}
\end{eqnarray}

By substituting the expression for $ x_v + x_u $ into the expression defining $ \bar{\kappa}(t = \infty) $, we obtain, after some manipulation,
\begin{equation}
\frac{(x_{vv} + \alpha x_{vu} - \phi_{rs})^2}{\phi_{rs} (x_{vv} + \alpha x_{vu} + \phi_{rs})^2} = \frac{1}{2} \frac{\kappa_{vv}}{\gamma \rho}
\end{equation}
where $ \phi_{rs} \equiv \bar{\kappa}(t = \infty)/\kappa_{vv} $.  Here, $ rs $ stands for ``random sexual".

The steady-state equations for $ x_{vv} $ and $ x_{vu} $ give,
\begin{eqnarray}
&  &
(1 + \phi_{rs}) x_{vv} = A_{vv} (\frac{2 x_{vv} + \alpha x_{vu}}{2 (x_{vv} + \alpha x_{vu})})^2 (\phi_{rs} + x_{vv} + \alpha x_{vu})
\nonumber \\
&  &
(\alpha + \phi_{rs}) x_{vu} = B_{vv} (\frac{2 x_{vv} + \alpha x_{vu}}{2 (x_{vv} + \alpha x_{vu})})^2 (\phi_{rs} + x_{vv} + \alpha x_{vu}) 
\nonumber \\
&  &
+ B_{vu} \frac{(2 x_{vv} + \alpha x_{vu}) \alpha x_{vu}}{4 (x_{vv} + \alpha x_{vu})^2} (\phi_{rs} + x_{vv} + \alpha x_{vu})
\end{eqnarray}

Now, from Eq. (16) it can be seen that $ \phi_{rs} \leq x_{vv} + \alpha x_{vu} $, and so from Eq. (20) we have,
\begin{eqnarray}
&  &
\frac{x_{vv} + \alpha x_{vu} - \phi_{rs}}{x_{vv} + \alpha x_{vu} + \phi_{rs}} = \sqrt{\frac{\kappa_{vv}}{2 \gamma \rho} \phi_{rs}}
\nonumber \\
&  &
\Rightarrow
x_{vv} + \alpha x_{vu} = \phi_{rs} f(\phi_{rs}, \lambda)
\end{eqnarray}
where $ \lambda \equiv \kappa_{vv}/(2 \gamma \rho) $, $ f(\phi_{rs}, \lambda) \equiv
(1 + \sqrt{\lambda \phi_{rs}})/(1 - \sqrt{\lambda \phi_{rs}}) $. 

We now have,
\begin{eqnarray}
&  &
(1 + \phi_{rs}) x_{vv} = A_{vv} (1 + f(\phi_{rs}, \lambda)) \frac{(x_{vv} + \phi_{rs} f(\phi_{rs}, \lambda))^2}{4 \phi_{rs} f(\phi_{rs}, \lambda)^2}
\nonumber \\
&  &
(\alpha + \phi_{rs}) \frac{\phi_{rs} f(\phi_{rs}, \lambda) - x_{vv}}{\alpha} = B_{vv} (1 + f(\phi_{rs}, \lambda))
\times
\nonumber \\
&  &
 \frac{(x_{vv} + \phi_{rs} f(\phi_{rs}, \lambda))^2}{4 \phi_{rs} f(\phi_{rs}, \lambda)^2} 
\nonumber \\
&  &
+ B_{vu} (1 + f(\phi_{rs}, \lambda)) 
\times 
\nonumber \\
&  &
\frac{(\phi_{rs} f(\phi_{rs}, \lambda) + x_{vv}) (\phi_{rs} f(\phi_{rs}, \lambda) - x_{vv})}{4 \phi_{rs} f(\phi_{rs}, \lambda)^2}
\end{eqnarray}

The second equation gives,
\begin{eqnarray}
&  &
(\alpha + \phi_{rs}) \phi_{rs} f(\phi_{rs}, \lambda) - (\alpha + \phi_{rs}) x_{vv} =
\nonumber \\
&  &
\alpha \frac{1 + f(\phi_{rs}, \lambda)}{4 \phi_{rs} f(\phi_{rs}, \lambda)^2} 
\times
\nonumber \\
&  &
[B_{vv} (\phi_{rs} f(\phi_{rs}, \lambda) + x_{vv})^2 + B_{vu} (\phi_{rs}^2 f(\phi_{rs}, \lambda)^2 - x_{vv}^2)]
\nonumber \\
\end{eqnarray}

Now, $ \phi_{rs}^2 f(\phi_{rs}, \lambda)^2 - x_{vv}^2 = -(\phi_{rs} f(\phi_{rs}, \lambda) + x_{vv})^2 + 2 \phi_{rs}^2 f(\phi_{rs}, \lambda)^2 + 2\phi_{rs} f(\phi_{rs}, \lambda) x_{vv} $, and so, after some manipulation, we obtain,
\begin{eqnarray}
&  &
x_{vv} = \phi_{rs} f(\phi_{rs}, \lambda) 
\times 
\nonumber \\
&  &
[(\alpha + \phi_{rs}) f(\phi_{rs}, \lambda) - \frac{1}{2} \alpha B_{vu} (1 + f(\phi_{rs}, \lambda))]/
\nonumber \\
&  &
[\phi_{rs} f(\phi_{rs}, \lambda) (1 - 2 \alpha) 
\nonumber \\
&  &
+ \alpha (\frac{1}{2} 
B_{vu} (1 + f(\phi_{rs}, \lambda)) - f(\phi_{rs}, \lambda))]
\nonumber \\
\end{eqnarray}
where we made use of the fact that $ B_{vv} - B_{vu} =  -2 A_{vv} $ and $ 2 (1 + \phi_{rs}) x_{vv} = 
(1 + f(\phi_{rs}, \lambda))/(2 \phi_{rs} f(\phi_{rs}, \lambda)^2) A_{vv} (\phi_{rs} f(\phi_{rs}, \lambda) + x_{vv})^2 $.

Plugging the value of $ x_{vv} $ back into the first equation from Eq. (23) we obtain, after tedious algebra,
\begin{eqnarray}
&  &
[1 + \phi_{rs}]
\times 
\nonumber \\ 
&  &
[\phi_{rs} f(\phi_{rs}, \lambda) - \alpha(\frac{1}{2} B_{vu} (1 + f(\phi_{rs}, \lambda)) - f(\phi_{rs}, \lambda))] 
\times \nonumber \\
&  &
[\phi_{rs} f(\phi_{rs}, \lambda) (1 - 2 \alpha) + \alpha (\frac{1}{2} B_{vu} (1 + f(\phi_{rs}, \lambda)) 
- f(\phi_{rs}, \lambda))]
\nonumber \\
&  &
= A_{vv} (1 - \alpha)^2 f(\phi_{rs}, \lambda) (1 + f(\phi_{rs}, \lambda)) \phi_{rs}^2
\end{eqnarray}

Now, $ 1 + f(\phi_{rs}, \lambda) = 2/(1 - \sqrt{\lambda \phi_{rs}}) $, and so, multiplying both sides by
$ (1 - \sqrt{\lambda \phi_{rs}})^2 $ gives,
\begin{equation}
\Lambda_1 \lambda \phi_{rs} + \Lambda_2 \sqrt{\lambda \phi_{rs}} - \Lambda_3 = 0
\end{equation}
where,
\begin{eqnarray}
&  &
\Lambda_1 \equiv (\phi_{rs} + 1) (\phi_{rs} + \alpha) (\phi_{rs} (1 - 2 \alpha) - \alpha)
\nonumber \\
&  &
\Lambda_2 \equiv 2 (\phi_{rs} + 1) ((1 - 2 \alpha) \phi_{rs}^2 - \alpha^2 (1 - p) \phi_{rs} + \alpha^2 p) 
\nonumber \\
&  &
- 2 A_{vv} (1 - \alpha)^2 \phi_{rs}^2
\nonumber \\
&  &
\Lambda_3 \equiv 2 A_{vv} (1 - \alpha)^2 \phi_{rs}^2 
\nonumber \\
&  &
- (\phi_{rs} + 1)(\phi_{rs} - \alpha p) (\phi_{rs} (1 - 2 \alpha) + \alpha p)
\nonumber \\
\end{eqnarray}

The solution that is chosen is the one that gives $ \lambda \rightarrow 0 $ as $ \Lambda_3 \rightarrow 0 $, since $ \Lambda_3 = 0 $ is the equation defining the steady-state mean fitness when there is no cost for sex.

\section{Comparison of the Asexual and Sexual Replication Strategies}

We now compare the various replication strategies.  We consider first the case where there is no cost for sex, so that $ \kappa_{vv}/(\gamma \rho) = 0 $, followed by the case where there is a non-zero cost for sex, so that $ \kappa_{vv}/(\gamma \rho) > 0 $.

The ratio $ \kappa_{vv}/(\gamma \rho) $ measures the cost for sex because it may be interpreted as the ratio of the characteristic time a haploid spends looking for another haploid with which to fuse, which is on the order of $ 1/(\gamma \rho) $, to the characteristic time it takes newly formed diploid cell to grow to maturity and divide, which is on the order of $ 1/\kappa_{vv} $.  When this ratio is small, then the fraction of the organism's life cycle that is devoted to the haploid fusion process is small, so that the time cost associated with sex is small.  Conversely, when this ratio is large, then the time cost associated with sex is large as well.

\subsection{Case $ 1 $:  $ \kappa_{vv}/(\gamma \rho) = 0 $}

When $ \kappa_{vv}/(\gamma \rho) = 0 $, the normalized mean fitness for the population replicating with the selective sexual replication strategy is given by,
\begin{equation}
\phi_{ss}^2 - (2 A_{vv} (1 - \alpha) - 1 + \alpha p) \phi_{ss} - \alpha p = 0
\end{equation}
while the normalized mean fitness for the population replicating with the random sexual replication strategy is given by,
\begin{equation}
2 A_{vv} (1 - \alpha)^2 \phi_{rs}^2 - (\phi_{rs} + 1) (\phi_{rs} - \alpha p) (\phi_{rs} (1 - 2 \alpha) + \alpha p) = 0
\end{equation}

The central result of this subsection is that $ \phi_{ss} > \phi_{rs} > \phi_{a} $, except when $ p = 0 $, $ p = 1 $, $ \alpha = 0 $, or $ \alpha = 1 $, in which case $ \phi_{ss} = \phi_{rs} = \phi_{a} $.  

We will prove this result in two steps:  First we will prove that $ \phi_{ss} = \phi_{rs} = \phi_{a} $ for $ p = 0 $, $ p = 1 $, $ \alpha = 0 $, or $ \alpha = 1 $.  We will then prove that $ \phi_{ss} > \phi_{rs} > \phi_{a} $ as long as $ \alpha, p \in (0, 1) $.

\subsubsection{Proof that $ \phi_{ss} = \phi_{rs} = \phi_{a} $ at $ p = 0, 1 $ and/or $ \alpha = 0, 1 $}

When $ p = 0 $, $ \phi_{ss} $, $ \phi_{rs} $ are obtained by solving,
\begin{eqnarray}
&  &
\phi_{ss} (\phi_{ss} + 1 - (1 - \alpha) s) = 0
\nonumber \\
&  &
\phi_{rs}^2 (s (1 - \alpha)^2 - (1 - 2 \alpha) (\phi_{rs} + 1)) = 0
\end{eqnarray}
which have the solutions $ \phi_{ss} = \phi_{sr} = 0 $.  We choose these solutions, because they are the ones that are physical.

When $ p = 1 $, $ \phi_{ss} $, $ \phi_{rs} $ are obtained by solving,
\begin{eqnarray}
&  &
(\phi_{ss} - 1) (\phi_{ss} + \alpha) = 0
\nonumber \\
&  &
(\phi_{sr} - 1) (\phi_{sr}^2 (1 - 2 \alpha) + \alpha^2) = 0
\end{eqnarray}
so that $ \phi_{ss} = \phi_{sr} = 1 $.

When $ \alpha = 0 $, we obtain,
\begin{eqnarray}
&  &
\phi_{ss} (\phi_{ss} - (2 A_{vv} - 1)) = 0
\nonumber \\
&  &
\phi_{rs}^2 (\phi_{rs} - (2 A_{vv} - 1)) = 0
\end{eqnarray}
These two equations both admit the solutions $ 0 $ and $ 2 A_{vv} - 1 $.  Since $ \phi_{ss} = \phi_{rs} = 1 $ for $ p = 1 $, and $ \phi_{ss} = \phi_{rs} = 0 $ for $ p = 0 $, by continuity it follows that $ \phi_{ss} = \phi_{rs} = 2 A_{vv} - 1 $ for $ p \in [p_{crit}, 1] $, and $ 0 $ for $ p \in [0, p_{crit}] $.  

When $ \alpha = 1 $, we obtain,
\begin{eqnarray}
&  &
(\phi_{ss} - p) (\phi_{ss} + 1) = 0
\nonumber \\
&  &
(\phi_{rs} - p)^2 (\phi_{rs} + 1) = 0
\end{eqnarray}
so that $ \phi_{ss} = \phi_{rs} = p $.

Note that in all cases, we have $ \phi_{ss} = \phi_{rs} = \phi_a $, as we wished to show.

\subsubsection{Proof that $ \phi_{ss} > \phi_{rs} > \phi_{a} $ when $ \alpha, p \in (0, 1) $}

When $ \alpha, p \in (0, 1) $, we claim that Eq. (30) has a solution in $ (\alpha p, p) $.  By continuity, we expect that this solution is the value of $ \phi_{rs} $ as a function of $ p $, since it is the solution that is consistent with $ \phi_{rs} = 0 $ at $ p = 0 $ and $ \phi_{rs} = 1 $ at $ p = 1 $.

When $ \phi_{rs} = \alpha p $, Eq. (30) evaluates to $ 2 A_{vv} (1 - \alpha)^2 \alpha^2 p^2 > 0 $.  When $ \phi_{rs} = p $, Eq. (30) evaluates to $ -(1 - \alpha)^2 p^2 (1 - p) (1 - s (1 - p)) < 0 $.

By the Intermediate Value Theorem, it follows that Eq. (30) has a solution in the interval $ (\alpha p, p) $.  This of course shows that $ \phi_{rs} > \alpha p $.  

We now claim that $ \phi_{rs} \neq 2 A_{vv} - 1 $.  For, if $ \phi_{rs} = 2 A_{vv} - 1 $, then from Eq. (30) we have,
\begin{equation}
(\phi_{rs} - p)^2 = 0
\end{equation}
which implies that $ \phi_{rs} = p $.  But this means that $ 2 p + s (1 - p)^2 - 1 = p \Rightarrow 
(1 - p) (1 - s (1 - p)) = 0 \Rightarrow p = 1 \Rightarrow\Leftarrow $, since $ p \in (0, 1) $ by assumption.  Therefore, $ \phi_{rs} \neq 2 A_{vv} - 1 $, as claimed.

If we can now show that there exists a $ p \in (0, 1) $ such that $ \phi_{rs} > 2 A_{vv} - 1 $, then we will have proven that $ phi_{rs} > 2 A_{vv} - 1 $ for all $ p \in (0, 1) $.  For otherwise, by the Intermediate Value Theorem we would be able to find a $ p \in (0, 1) $ such that $ \phi_{rs} = 2 A_{vv} - 1 \Rightarrow\Leftarrow $.  

Now, $ p_{crit} $ is defined by the equation $ 2 p_{crit} + s (1 - p_{crit})^2 - 1 = \alpha p_{crit} $, so that $ p_{crit} = 1 \Rightarrow \alpha = 1 $ and $ p_{crit} = 0 \Rightarrow s = 1 $.  Since we are assuming $ \alpha \in (0, 1) $, we either have $ s < 1 $ or $ s = 1 $.  If $ s < 1 $, then $ p_{crit} \in (0, 1) $, so since $ 2 A_{vv} - 1 = \alpha p $ at $ p = p_{crit} $, we have that $ \phi_{rs} > 2 A_{vv} - 1 $ at $ p = p_{crit} \in (0, 1) $, thereby proving that $ \phi_{rs} > 2 A_{vv} - 1 $ for $ p \in (0, 1) $.  Taking the limit $ s \rightarrow 1 $ gives that $ \phi_{rs} \geq 2 A_{vv} - 1 $ for $ p \in (0, 1) $, which of course implies that $ \phi_{rs} > 2 A_{vv} - 1 $ for $ p \in (0, 1) $, since $ \phi_{rs} \neq 2 A_{vv} - 1 $ for $ p \in (0, 1) $.

We have now shown that $ \phi_{rs} > \alpha p, 2 A_{vv} - 1 $ when $ \alpha, p \in (0, 1) $, and so $ \phi_{rs} > \phi_{a} $ for $ \alpha, p \in (0, 1) $.  We now turn to proving that $ \phi_{ss} > \phi_{rs} $ for $ \alpha, p \in (0, 1) $.

If we define $ \Omega = 2 A_{vv} (1 - \alpha) - 1 + \alpha p $, then $ \phi_{ss}^2 = \Omega \phi_{ss} + \alpha p $.  If $ \phi_{ss} = \phi_{sr} $ for some $ p \in (0, 1) $, then defining $ \phi = \phi_{ss} = \phi_{sr} $, we obtain from Eq. (30) that,
\begin{eqnarray}
&  &
2 A_{vv} (1 - \alpha)^2 (\Omega \phi + \alpha p) 
\nonumber \\
&  &
= \phi (\Omega + 1 - \alpha p)(\phi (1 - 2 \alpha) + \alpha p)
\nonumber \\
&  &
= (\Omega + 1 - \alpha p)[((1 - 2 \alpha) \Omega + \alpha p) \phi + (1 - 2 \alpha) \alpha p]
\nonumber \\
&  &
\Rightarrow
[2 A_{vv} (1 - \alpha)^2 \Omega - (\Omega + 1 - \alpha p)((1 - 2 \alpha) \Omega + \alpha p)] \phi
\nonumber \\
&  &
= (\Omega + 1 - \alpha p) (1 - 2 \alpha) \alpha p - 2 A_{vv} (1 - \alpha)^2 \alpha p
\end{eqnarray}

Now, $ \Omega + 1 - \alpha p = 2 A_{vv} (1 - \alpha) $, so that,
\begin{equation}
[\Omega - p] \phi = -\alpha p
\end{equation}

So, multiplying Eq. (29) by $ (\Omega - p)^2 $ gives,
\begin{eqnarray}
&  &
\alpha^2 p^2 + \Omega (\Omega - p) \alpha p - (\Omega - p)^2 \alpha p = 0 
\nonumber \\
&  &
\Rightarrow
- (1 - \alpha) (1 - p) (1 - s (1 - p)) = 0
\end{eqnarray}
which is impossible for $ \alpha, p \in (0, 1) $.  Therefore, $ \phi_{ss} \neq \phi_{rs} $ for $ p \in (0, 1) $.

Now, let us look at $ d \phi_{ss}/dp $ and $ d \phi_{rs}/dp $ at $ p = 1 $.  For $ \phi_{ss} $ we have,
\begin{equation}
(\frac{d \phi_{ss}}{d p})_{p = 1} = \frac{2}{1 + \alpha}
\end{equation}
while for $ \phi_{rs} $ we have,
\begin{equation}
(\frac{d \phi_{sr}}{d p})_{p = 1} = 2
\end{equation}

Therefore, for $ p $ near $ 1 $, we have,
\begin{eqnarray}
&  &
\phi_{ss} = 1 - \frac{2}{1 + \alpha} (1 - p)
\nonumber \\
&  &
\phi_{rs} = 1 - 2 (1 - p)
\end{eqnarray}
so since $ 2 > 2/(1 + \alpha) $, it follows that $ \phi_{ss} > \phi_{rs} $ for $ p < 1 $, but $ p $ close to $ 1 $.
But then $ \phi_{ss} > \phi_{rs} $ for all $ p \in (0, 1) $, for otherwise there exists a $ p \in (0, 1) $ for which $ \phi_{ss} \leq \phi_{rs} $, which implies by the Intermediate Value Theorem that there exists a $ p \in (0, 1) $ such that $ \phi_{ss} = \phi_{rs} \Rightarrow\Leftarrow $.

We have now established that $ \phi_{ss} > \phi_{sr} > \phi_{a} $ for $ \alpha, p \in (0, 1) $, as we set out to prove.

\subsection{Case $ 2 $:  $ \kappa_{vv}/(\gamma \rho) > 0 $}

The central result of this subsection is that $ \phi_{ss} > \phi_{rs} $ for $ \alpha, p \in (0, 1) $, with equality only occurring when $ p = 0 $, $ p = 1 $, $ \alpha = 0 $, or $ \alpha = 1 $.

Suppose we can show that $ \phi_{ss} \neq \phi_{rs} $ for $ \alpha, p \in (0, 1) $, independently of the value of $ \kappa_{vv}/(\gamma \rho) $.  Then if $ \phi_{ss} \leq \phi_{rs} $ for some $ \alpha, p \in (0, 1) $ and for some value of $ \kappa_{vv}/(\gamma \rho) \geq 0 $, it follows from continuity and the Intermediate Value Theorem that $ \phi_{ss} = \phi_{rs} $ for some value of $ \kappa_{vv}/(\gamma \rho) \geq 0 $, since $ \phi_{ss} > \phi_{rs}  $ for $ \alpha, p \in (0, 1) $ when $ \kappa_{vv}/(\gamma \rho) = 0 $.  This is a contradiction, and so we must have that $ \phi_{ss} > \phi_{rs} $ for all $ \alpha, p \in (0, 1) $ and $ \kappa_{vv}/(\gamma \rho) \geq 0 $.

Therefore, to prove the central result of this subsection, we will show that $ \phi_{ss} = \phi_{rs} $ for $ p = 0, 1 $ and/or $ \alpha = 0, 1 $, and that $ \phi_{ss} \neq \phi_{rs} $ for $ \alpha, p \in (0, 1) $.

Working again with the definition $ \lambda = \kappa_{vv}/(2 \gamma \rho) $, we have from Eq. (14) that,
\begin{equation}
\frac{(\phi_{ss}^2 - (2 A_{vv} (1 - \alpha) - 1 + \alpha p) \phi_{ss} - \alpha p)^2}{(\phi_{ss} + 1)^2 (\phi_{ss} + \alpha)^2} = \lambda \phi_{ss}
\end{equation}

As $ \lambda $ increases from $ 0 $ to $ \infty $, we expect $ \phi_{ss} $ to decrease from its maximal value down to $ 0 $.  In this regime, we would like to determine the sign of the expression $ \phi_{ss}^2 - (2 A_{vv} (1 - \alpha) - 1 + \alpha p) \phi_{ss} - \alpha p  $.  

Consider the polynomial $ x^2 - A x - B $, which has the roots $ (1/2) [A \pm \sqrt{A^2 + 4 B}] $.  If $ B > 0 $, then the ``+" root, denoted $ r_+ $, is positive, and the ``-" root, denoted $ r_- $, is negative.  Therefore, writing $ x^2 - A x - B = (x - r_{-}) (x - r_{+}) $, we may note that when $ x \in [0, r_{+}] $, $ x - r_- \geq 0 $ and $ x - r_+ \leq 0 $, so that $ x^2 - A x - B \leq 0 $.

The positive root of $ \phi_{ss}^2 - (2 A_{vv} (1 - \alpha) - 1 + \alpha p) \phi_{ss} - \alpha p $ is the value of $ \phi_{ss} $ when $ \lambda = 0 $.  Since $ \phi_{ss} $ then decreases to $ 0 $ as $ \lambda $ increases, it follows from our analysis that $ \phi_{ss}^2 - (2 A_{vv} (1 - \alpha) - 1 + \alpha p) \phi_{ss} - \alpha p $ is zero or negative for $ \lambda \geq 0 $.  Therefore, taking the square roots of both sides of Eq. (42) gives,
\begin{equation}
\frac{\alpha p + (2 A_{vv} (1 - \alpha) - 1 + \alpha p) \phi_{ss} - \phi_{ss}^2}{(\phi_{ss} + 1)(\phi_{ss} + \alpha)} = \sqrt{\lambda \phi_{ss}}
\end{equation}

If $ \phi_{ss} = \phi_{rs} $ for a given $ \alpha, p \in [0, 1] $ and $ \lambda \geq 0 $, then we may insert the expression above into Eq. (27) defining $ \phi_{rs} $.  If we set $ \phi = \phi_{ss} = \phi_{rs} $, then we obtain, after some manipulation,
\begin{eqnarray}
&  &
0 = 
[\phi (1 - 2 \alpha) - \alpha] 
\times \nonumber \\
&  &
[\phi^2 - (2 A_{vv} (1 - \alpha) - 1 + \alpha p) \phi - \alpha p]^2
\nonumber \\
&  &
- 2 [\phi + 1] 
\times \nonumber \\
&  &
[(1 - 2 \alpha) \phi^2 - \alpha^2 (1 - p) \phi + \alpha^2 p] 
\times \nonumber \\
&  &
[\phi^2 - (2 A_{vv} (1 - \alpha) - 1 + \alpha p) \phi - \alpha p]
\nonumber \\
&  &
 + 2 A_{vv} (1 - \alpha)^2 \phi^2 [\phi^2 - (2 A_{vv} (1 - \alpha) - 1 + \alpha p) \phi - \alpha p]
\nonumber \\
&  &
- 2 A_{vv} (1 - \alpha)^2 \phi^2 [\phi + 1] [\phi + \alpha] 
\nonumber \\
&  &
+ [\phi + 1]^2 [\phi + \alpha] [\phi - \alpha p][\phi (1 - 2 \alpha) + \alpha p]
= 0
\end{eqnarray}

This expression may be simplified (with the aid of a symbolic math package if necessary) to give,
\begin{eqnarray}
0 
& = & 
\alpha (1 - \alpha)^2 (1 - p) \phi_s^2 (1 - \phi_s) 
\times \nonumber \\
&  &
[r^2 p^3 + 3 r (1 - r) p^2 + (2 (1 - r)^2 + r^2) p + r (1 - r)]
\nonumber \\
\end{eqnarray}
which implies that either $ \alpha = 0, 1 $, $ p = 1 $, or $ \phi = 0, 1 $.  Since $ \phi = 0 $ is equivalent to $ p = 0 $ when $ \alpha > 0 $, and since $ \phi = 1 $ is equivalent to $ p = 1 $ and $ \lambda = 0 $, we obtain that $ \phi_{ss} = \phi_{rs} $ only when $ \alpha = 0, 1 $ or $ p = 0, 1 $.  If $ \alpha, p \in (0, 1) $, then $ \phi_{ss} \neq \phi_{rs} $.

Therefore, we have proven that $ \phi_{ss} = \phi_{rs} $ when $ p = 0 $, $ p = 1 $, $ \alpha = 0 $, or $ \alpha = 1 $, and $ \phi_{ss} > \phi_{rs} $ for $ \alpha, p \in (0, 1) $, independently of the value of $ \kappa_{vv}/(\gamma \rho) $.

\section{Discussion}

The two key results of this paper are that a sexual population employing a random mating strategy will outcompete an asexual population when the cost for sex is negligible, and that a sexual population using a selective mating strategy will outcompete a sexual population using a random mating strategy.
The only exceptions are the boundary cases $ p = 0, 1 $ and $ \alpha = 0, 1 $.  However, even here, the mean fitnesses of the two sexual strategies are identical.  Furthermore, when the cost for sex is negligible, then the mean fitnesses of the sexual strategies are identical to that of the asexual strategy.

That random mating provides a selective advantage over asexual replication is an interesting result, because the strategy can lead to the formation of diploids with completely defective genomes.  Presumably, however, the fitness benefit provided by the formation of diploids with two functional chromosomes outweighs the fitness cost associated with the formation of diploids with two defective chromosomes, leading to an overall advantage for the strategy.  Nevertheless, because the selective mating strategy does not produce genomes with defective chromosomes, this strategy has an advantage over the random mating strategy.

When the cost for sex is negligible, an analysis of the mean fitnesses $ \phi_a $, $ \phi_{ss} $, and $ \phi_{rs} $ near $ p = 1 $ yields some interesting results.  We have $ \phi_a = 2 A_{vv} - 1 = 1 - 2 (1 - p) + s (1 - p)^2 $.  We have also shown that when $ p $ is close to $ 1 $, then to first-order in $ 1 - p $ we have $ \phi_{rs} = 1 - 2 (1 - p) $ and $ \phi_{ss} = 1 - (2/(1 + \alpha)) (1 - p) $.  Therefore, we may note that the random mating strategy and the asexual strategy are identical to first-order in $ 1 - p $, while the selective mating strategy already outcompetes both the random and asexual strategies.

However, from Eq. (30) we obtain,
\begin{equation}
(\frac{d^2 \phi_{rs}}{d p^2})_{p = 1} = 2 (s + 2 (\frac{\alpha}{1 - \alpha})^2)
\end{equation}
and so, to second-order in $ 1 - p $ we have that,
\begin{equation}
\phi_{rs} = 1 - 2 (1 - p) + [s + 2 (\frac{\alpha}{1 - \alpha})^2] (1 - p)^2
\end{equation}
Comparing the second-order expression for $ \phi_{rs} $ to $ \phi_a $, we see that $ \phi_{rs} $ exceeds $ \phi_a $ by $ 2 (\alpha/(1 - \alpha))^2 (1 - p)^2 $ when $ p $ is close to $ 1 $.

Figure 3 shows a plot of the two sexual replication strategies when there is no cost for sex and the asexual strategy.

When there is a cost for sex, then when $ p = 1 $ the asexual population outcompetes both sexual populations.  This makes sense, for when $ p = 1 $ replication is error-free, and hence at steady-state the asexual population consists only of the wild-type.  In this case, genetic recombination will not improve fitness, since there are no defective chromosomes in the population to begin with.

However, because sexual replication when there is no cost for sex will outcompete an asexual replication when $ p \in (0, 1) $, it follows that if the cost for sex is sufficiently low, then below a certain value of $ p $ a sexual population will outcompete the asexual population.  Presumably, the higher the cost for sex, the smaller $ p $ must be before the selective advantage for sexual replication is sufficiently large to outweigh the cost.  

This behavior only persists up to a maximal cost for sex, beyond which asexual replication outcompetes sexual replication at all replication fidelities.  The reason for this is that the sexual and asexual mean fitnesses converge to $ 0 $ as $ p \rightarrow 0 $.  As a result, once the cost for sex is sufficiently high, the fitness advantage of the sexual strategy for the values of $ p $ where the sexual strategy can outcompete the asexual strategy is too small to overcome the cost for sex.

Indeed, from Eq. (14) it may be shown that $ (d \phi_{ss}/d p)_{p = 0} = 0 $ when $ \kappa_{vv}/(\gamma \rho) > 0 $.  Since $ (d \phi_a/d p)_{p = 0} = \alpha $, it follows for $ \alpha \in (0, 1) $ that $ \phi_{ss} < \phi_a $ in a neighborhood of $ p $ sufficiently close to $ 0 $.  Since $ \phi_{rs} \leq \phi_{ss} $, the same condition holds for $ \phi_{rs} $ as well.  However, if any of the sexual populations can outcompete the asexual population when $ p $ is sufficiently large, then there must exist another value of $ p $ below which asexual replication outcompetes sexual replication, and above which sexual replication outcompetes asexual replication.

The complete picture is then one where, for a non-zero cost for sex, the asexual population outcompetes a sexual population above a certain replication fidelity.  Below this replication fidelity, the sexual population outcompetes the asexual population.  Finally, once the replication fidelity becomes sufficiently low, the asexual population again outcompetes the sexual population.  

Based on this analysis, we expect that, as the cost for sex increases from $ 0 $ to $ \infty $, the region of replication fidelities where sexual replication outcompetes asexual replication starts at $ (0, 1) $, gradually shrinks, and eventually disappears once the cost for sex crosses a threshold value.  Of course, because the random mating strategy has a lower fitness than the selective mating strategy, the region of replication fidelities where the random mating strategy outcompetes asexual replication is a subset of the region of replication fidelities where the selective mating strategy outcompetes asexual replication.

Figure 4 shows a plot of the asexual and sexual mean fitnesses when there is a non-zero cost for sex.  Figure 5 shows a plot of the regions for the selective advantages of the various replication strategies as a function of $ \kappa_{vv}/(\gamma \rho) $.

\begin{figure}
\includegraphics[width = 0.9\linewidth, angle = 0]{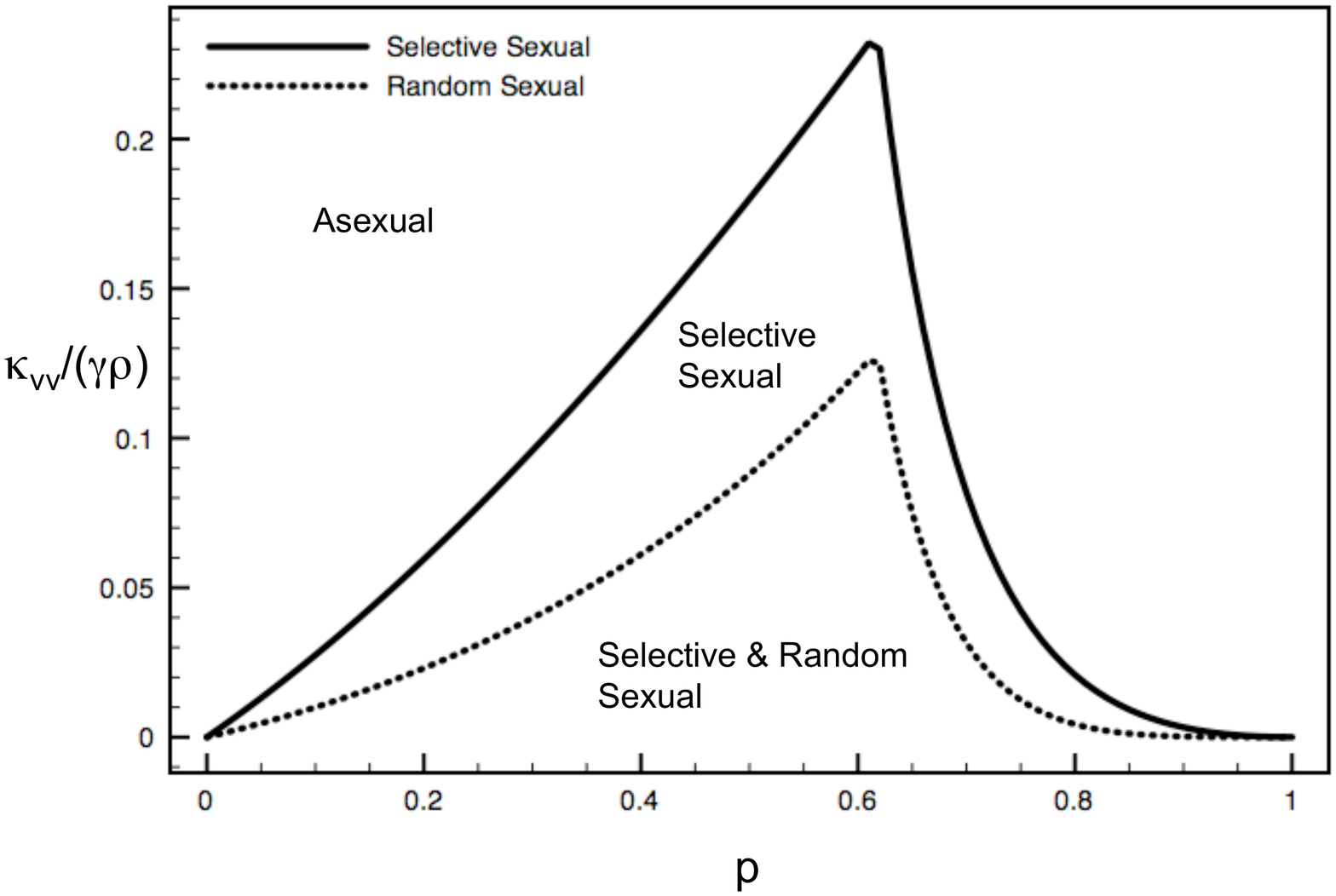}
\caption{Regimes where the asexual, selective sexual, and random sexual replication strategies are respectively advantageous, as a function of $ p $ and $ \kappa_{vv}/(\gamma \rho) $.  The region below the dotted curve is the region where both the random and selective mating strategies outcompete the asexual strategy.  The region below the solid curve and above the dotted curve is the region where only the selective mating strategy outcompetes the asexual strategy.  The region above the solid curve is the region where the asexual strategy dominates.  Note that as the cost for sex increases, the regions where the sexual strategies are advantageous shrink and eventually disappear.  The random mating strategy ceases to be advantageous at a lower cost than the selective strategy.  The parameters chosen are $ \alpha = r = 1/2 $.}
\end{figure}

Although we have shown that a selective mating strategy will outcompete a random strategy, this analysis is based on the assumption that a selective mating strategy is not inherently slower than a random mating strategy.  As discussed in the Introduction, this is in general not true, since there is a time cost associated with determining the genome of a potential haploid mate.  This time cost leads to an additional fitness cost associated with a selective mating strategy that is not incurred by a random mating strategy.  When the replication fidelity is either sufficiently high or sufficiently low, then the benefits of a selective mating strategy are not sufficient to overcome the fitness costs, and so the random mating strategy dominates.  However, at intermediate replication fidelities, the selective mating strategy may indeed outcompete the random mating strategy, assuming that the fitness cost associated with selective mating is not too high.  Once the fitness cost associated with selective mating becomes sufficiently high, then the random mating strategy may outcompete the selective mating strategy at all replication fidelities. 

In summary, when the time cost for a selective mating strategy is taken into account, an analysis of the regimes where the selective and random mating strategies are expected to be respectively dominant may produce a curve that is analogous to the ones in Figure 5.

\section{Conclusions and Future Research}

This paper developed a set of simplified models describing asexual and sexual replication in a unicellular population consisting of two-chromosomed, diploid genomes.  We considered two types of sexual replication strategies:  A selective mating strategy, where only viable haploids are allowed to fuse, and a random mating strategy, where all haploids are allowed to participate in the replication process.  We assumed that haploid fusion was a second-order rate process.

We found that, when the cost for sex is negligible, both the selective and random mating strategies lead to a greater mean fitness than asexual replication.  Nevertheless, we found that the selective mating strategy has a higher mean fitness than the random mating strategy, as long as the additional fitness penalty associated with a selective mating strategy is negligible.

Further analysis suggested that sexual replication for both the selective and random mating strategies is favored at intermediate mutation rates, and when the cost for sex is sufficiently low.  Once the cost for sex becomes sufficiently high, the selective advantage for sexual replication disappears entirely.  

The results of this paper therefore suggest that sex is favored in slowly replicating organisms and high population densities.  While this is consistent with previous studies \cite{Tannenbaum:06, TannFon:06, LeeTann:07, Tannenbaum:07}, what is interesting is that this result holds even with a random mating strategy.

For future work, we will consider models where it is not necessarily true that $ \kappa_{uu} = 0 $, but rather we will allow for positive values of $ \kappa_{uu} $.  This will make our model more consistent with quasispecies models that often assume a small, but positive growth rate for organisms with defective genomes.  

Furthermore, although our model provides results that are broadly consistent with actual organismal behavior, we have nevertheless worked with a highly simplified sexual replication model that does not exactly correspond to the sexual replication pathway employed by unicellular organisms.  

In our model, a mature diploid divides into two haploids, the haploids fuse, and then the resulting diploid divides into two cells.  Thus, the diploid mitosis occurs {\it after} haploid fusion.  By contrast, in {\it Saccharomyces cerevisiae} (Baker's yeast), for example, a mature diploid divides into two diploids, which then divide into  four haploids.  The haploids then fuse, and the resulting diploids grow to maturity to begin the process again.  Thus, for yeast, diploid mitosis occurs {\it before} haploid fusion.  We suspect that the relative positions of the diploid mitosis and haploid fusion stages in a sexual replication cycle can affect the mutation-selection balance, and so we would like to develop more realistic models corresponding to the sexual replication pathways in actual unicellular organisms.  We should emphasize, however, that we believe that the central results of this paper will hold even when we consider a somewhat different sexual replication pathway.

Finally, one of the simplifying assumptions made in this paper is that the organisms have genomes consisting of two chromosomes.  For future research, we plan to study genomes consisting of arbitrary numbers of chromosomes.  Within this context, we plan to consider more complex fitness landscapes and the role of intra-genomic recombination.   

\begin{acknowledgments}

This research was supported by a Start-Up Grant from the United States -- Israel Binational Science Foundation, and by an Alon Fellowship from the Israel Science Foundation.  The author would also like to thank J.F. Fontanari for helpful conversations leading to the completion of this work.

\end{acknowledgments}

\end{document}